\documentclass[a4paper]{article}

\usepackage{INTERSPEECH2021}
\usepackage{hyperref}
\usepackage{array}
\newcolumntype{?}{!{\vrule width 1pt}}
\usepackage{multirow}
\usepackage{footnote}
\makesavenoteenv{tabular}
\makesavenoteenv{table}

\title{Evaluating the  \textbf{C}OVID-19 \textbf{Ide}ntification \textbf{R}esNet  (\textbf{CIdeR}) on the INTERSPEECH COVID-19 from Audio Challenges}
\name{Alican Akman*$^1$\thanks{*Equal contribution}, Harry Coppock*$^1$, Alexander Gaskell$^1$, Panagiotis Tzirakis$^1$, Lyn Jones$^2$, \\Björn W.\ Schuller$^{1,3}$}
\address{
  $^1$GLAM -- Group on Language, Audio, \& Music, Imperial College London, London, UK\\ 
  $^2$Radiology Department, North Bristol NHS Trust, Bristol, UK\\
$^3$Chair of Embedded Intelligence for Health Care and Wellbeing, University of Augsburg, Germany}

\email{harry.coppock@imperial.ac.uk, a.akman21@imperial.ac.uk}

\begin{document}
\maketitle
\begin{abstract}
 We report  on cross-running the recent COVID-19 Identification ResNet (CIdeR) on the two Interspeech 2021 COVID-19 diagnosis from cough and speech audio challenges: ComParE and DiCOVA. CIdeR is an end-to-end deep learning neural network originally designed to classify whether an individual is COVID-positive or COVID-negative based on coughing and breathing audio recordings from a published crowdsourced dataset. In the current study, we demonstrate the potential of CIdeR at binary COVID-19 diagnosis from both the COVID-19 Cough and Speech Sub-Challenges of INTERSPEECH 2021, ComParE and DiCOVA. CIdeR achieves significant improvements over several baselines. 
\end{abstract}

\section{Introduction}
The current coronavirus pandemic (COVID-19), caused by the  severe-acute-respiratory-syndrome-coronavirus 2 (SARS-CoV-2), has infected a confirmed 126 million people and resulted in 2,776,175 deaths (WHO)\footnote{as of 29th March 2021 https://www.who.int/emergencies/diseases/novel-coronavirus-2019}. Mass testing schemes offer the option to monitor and implement a selective isolation policy to control the pandemic without the need for regional or national lockdown\cite{Peto2020weeklymasstesting}. However, physical mass testing methods, such as the Lateral Flow Test (LFT) have come under criticism since the tests divert limited resources from more critical services \cite{Wisem4690, Holt2021} and due to suboptimal diagnostic accuracy. Sensitivities of 58\,\% have been reported for self-administered LFTs \cite{Mahasem4469}, unacceptably low when used to detect active virus, a context where high sensitivity is essential to prevent the reintegration into society of falsely reassured infected test recipients \cite{Moysen90}.

 Investigating the potential for digital mass testing methods is an alternative approach, based on findings that suggest a biological basis for identifiable vocal biomarkers caused by SARS-CoV-2's effects on the lower respiratory track \cite{Quatieri20}. This has recently been backed up by empirical evidence \cite{bartlpokorny2020voice}. Efforts have been made to collect and classify a range of different modality audio recordings of COVID-positive and COVID-negative individuals and several datasets have been released that use applications to collect the breath and cough of volunteer individuals. Examples include the ‘Coughvid’ \cite{orlandic2020coughvid}, ‘Breath for Science’\footnote{https://www.breatheforscience.com}, ‘Coswara’ \cite{SharmaKKRCRGG20}, COVID-19 sounds\footnote{https://www.covid-19-sounds.org/en/}, and ‘CoughAgainstCovid’ \cite{bagad2020cough}. In addition, to focus the attention of the audio processing community onto the task of binary classification of COVID-19 from audio, two INTERSPEECH competitions: the INTERSPEECH 2021 Computational Paralinguists Challenge (ComParE)\footnote{http://www.compare.openaudio.eu/}  \cite{schuller2021interspeech} with its COVID-19 Cough and Speech Sub-Challenges, and Diagnosing COVID-19 using acoustics (DiCOVA)\footnote{https://dicova2021.github.io/} \cite{muguli2021dicova} have been organised with this focus as their challenge.

Several studies have been published that propose machine learning-based COVID classifiers exploiting distinctive sound properties between positive and negative cases to classify these datasets. \cite{brown2020exploring} and \cite{ritwik2020covid} demonstrate that simple machine learning models perform well in these relatively small datasets. In addition, deep neural networks are exploited in \cite{laguarta2020covid,pinkas2020sars,imran2020ai4covid, minaDetecting} with proven performance at the COVID detection task. Although there are works that try to combine different modalities computing the representations separately, \cite{coppock2021end2end} (CIdeR) proposes an approach computing joint representation of a number of modalities. The  adaptability of this approach to different types of datasets has not to our knowledge been explored or reported. 

To this end, we present the results of the application of COVID-19 Identification ResNet (CIdeR), a recently developed end-to-end deep learning neural network optimised for binary COVID-19 diagnosis from cough and breath audio \cite{coppock2021end2end},  to the two COVID-19 cough and speech Challenges of INTERSPEECH 2021, ComParE and DiCOVA.

\section{Methods}

\subsection{Model} 
\begin{figure*}[ht]
\centering
    \includegraphics[width=14cm, height=5cm]{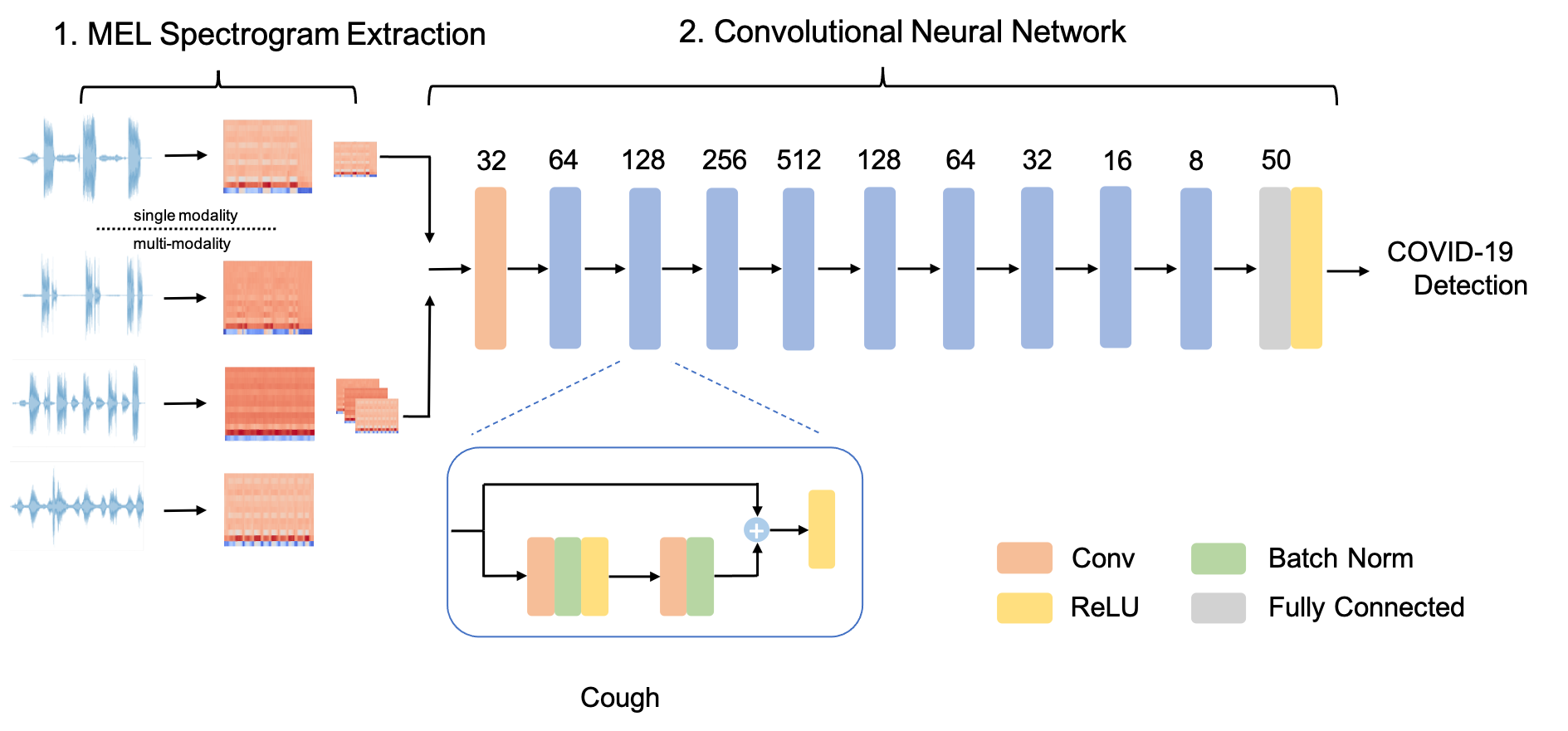}\caption{A schematic of the \textbf{C}OVID-19 \textbf{Ide}ntification \textbf{R}esNet, (\textbf{CIdeR}). The figure shows a blow-up of a residual block, consisting of convolutional, batch normalisation, and Rectified Linear Unit (ReLU) layers.}
\label{network_architecture}

\end{figure*}

CideR\cite{coppock2021end2end} is a 9 layer convolutional residual network. A schematic detailing of the model can be seen in Figure~\ref{network_architecture}. Each layer or block consists of a stack of convolutional layers with Rectified Linear Units (ReLUs). Batch normalisation\cite{batch_norm} also features in the residual units, acting as a source of regularisation and supporting training stability. A fully connected layer with sigmoid activation terminates the model yielding a single logit output which can be interpreted as an estimation of the probability of COVID-19. As detailed in Figure~\ref{network_architecture} the network is compatible with a varying number of modalities, for example,  if a participant has provided cough, deep breathing, and sustained vowel phonation audio recordings, they can be stacked in a depth wise manner and passed through the network as a single instance.

\subsection{Pre-processing} At training time, a window of s-seconds, which was fixed at 6 seconds for these challenges, is sampled from the audio recording randomly. If the audio recording is less than s-seconds long,  the sample is padded with repeated versions of itself. The sampled audio is then converted into Mel-Frequency Cepstral Coefficients (MFCCs) resulting in an image of width s * the sample rate and height equal to the number of MFCCs. Three data augmentation steps are then applied to the sample. First, the pitch of the recording is randomly shifted, secondly, bands of the Mel spectrogram are masked in the time and Mel coefficient axes and finally, Gaussian noise is added. At test time, the sampled audio recording is chunked into a set of s-second clips and processed in parallel. The mean of the set of logits is then returned as the final prediction.

\subsection{Baselines}

The DiCOVA team ran baseline experiments for the track 1 (coughing) sub-challenge; only the best performing (MLP) model's score was reported. For the track 2 (deep breathing/vowel phonation/counting) sub-challenge, however, baseline results were not provided. Baseline results were provided for the ComParE challenge but only Unweighted Average Recall (UAR) was reported rather than Area Under Curve of the Receiver Operating Characteristics curve (ROC-(AUC)). To allow comparison across challenges, we created new baseline results for the ComParE sub-challenges and the DiCOVA Track 2 sub-challenge, using the same baseline methods described for the DiCOVA Track 1 sub-challenge. The three baseline models applied to all four sub-challenge datasets were Logistic Regression (LR), Multi-layer Perceptron (MLP), and Random Forrest (RF), where the same hyperparameter configurations that were specified in the DiCOVA baseline algorithm was used \cite{muguli2021dicova}.

To provide a baseline comparison for the CIdeR track 2 results, we built a multimodal baseline model. We followed a similar strategy with the provided DiCOVA baseline algorithm, while extracting the features for each modality. Rather than individual training for different models, we developed an algorithm that concatenates input features from separate modalities. Then, this combined feature set was fed to the baseline models: LR, MLP, and RF.

We used 39 dimensional MFCCs as our feature type to represent the input sounds. For LR, we used Least Square Error (L2) as a penalty term. For MLP, we used a single hidden layer of size 25 with a Tanh activation layer and L2 regularisation. The Adam optimiser and a learning rate of 0.0001 was used. For RF, we built the model with 50 trees and split based on the gini impurity criterion.

\section{Datasets}

\subsection{ComParE}
ComParE hosted two COVID-19 related sub-challenges, the COVID Cough Sub-Challenge (CCS) and the COVID Speech Sub-Challenge (CSS). Both CCS and CSS are subsets of the crowd sourced Cambridge COVID-19 sound database \cite{brown2020exploring, han2021exploring}. CCS consists of 926 cough recordings from 397 participants. Participants provided 1-3 forced coughs resulting in a total of 1.63 hours of recording. CSS is made up of 893 recordings from 366 participants totalling 3.24 hours of recording. Participants were asked to recite the phrase \textit{``I hope my data can help manage the virus pandemic''} in their native language 1-3 times. The train-test splits for both sub-challenges are detailed in Table \ref{tab:compare}.

\begin{table}[th]
  \caption{ComParE sub-challenge dataset splits. Values specify the number of audio recordings, not the number of participants.}
  \label{tab:compare}
  \centering
  \begin{tabular}{ l | r | r | r | r | r | rl}
    \toprule
     & \multicolumn{3}{c}{CCS} &  \multicolumn{3}{c}{CSS}\\
    \cline{2-7}
    \# & train & val & test & train & val & test \\
    \hline
    COVID-postive & 71 & 48& 48 & 72& 142 & 94 \\
    COVID-negative& 215& 183& 183 & 243& 153 & 183 \\
    \hline
    Total& 286 & 231 & 208& 315& 295& 283 \\

    \bottomrule
  \end{tabular}

\end{table}

\begin{table}[th]
  \caption{DICOVA sub-challenge dataset splits. The test set labels were withheld by the DiCOVA team, contestants had to submit predictions for each test case, on which a final AUC was returned.}
  \label{tab:DiCOVA}
  \centering
  \begin{tabular}{ l | r | r | r | r }
    \toprule
     & \multicolumn{2}{c}{Track-1} &  \multicolumn{2}{c}{Track-2}\\
    \cline{2-5}
    \# & train + val & test & train + val & test \\
    \hline
    COVID-postive & 75 & blind & 60 & 21\\
    COVID-negative& 965 & blind & 930 & 188 \\
    \hline
    Total& 1,040 & 234 & 990 & 209 \\

    \bottomrule
  \end{tabular}

\end{table}

\begin{table*}[t]
  \caption{Results for CIdeR and a range of baseline models for 4 sub-challenges across the DiCOVA and ComParE challenges. Testing is performed on the held-out test fold once final model decisions have been made on the validation sets. The Area Under Curve of the Receiver Operating Characteristics curve (AUC(-ROC)) is displayed. A 95\,\% confidence interval is also shown following \cite{hanley_mcneil_1982}. CIdeR scores which are statistically higher than the best baseline results with a 95\,\% confidence are in bold. The three baseline models are Logistic Regression (LR), Multi-layer Perceptron (MLP), and Random Forrest (RF). All baseline models were trained on MFCC features.}
  
  \centering
  \begin{tabular}{ l | l | c | c | c | c}
    \toprule
    & sub-challenge*  & CIdeR & \multicolumn{3}{c}{Baseline} \\
    \cline{4-6}
    & & & LR & MLP & RF \\
    \midrule
    \multirow{2}{*}{DiCOVA} & Track 1** & $\textbf{.799} \pm .058$ &    -    & $.699 \pm .068$ & -\\
    & Track 2 & $.786 \pm .057$ & $.647 \pm .014$ & $.684 \pm .072$ & $.776 \pm .063$ \\
    \hline
    \multirow{2}{*}{ComParE} & CCS & $.732 \pm .068$ & $.722 \pm .069$ & $.765 \pm .065$ & $.753 \pm .066$ \\
    & CSS & $\textbf{.787} \pm .060$ & $.583 \pm .072$ & $.656 \pm .070$ & $.628 \pm .070$\\
    \bottomrule

  \end{tabular}
  
  \label{tab:results}
 
    \footnotesize{*Track 1: coughing, Track 2: deep breathing + vowel phonation + counting, CCS: coughing, CSS: speech --`\textit{ hope my data can help managethe  virus  pandemic}'}\\
    \footnotesize{** As the demographics were not provided for the Track 1 test set, when calculating the AUC confidence intervals, it was assumed that there was an equal number of COVID-positive and COVID-negative recordings. }
\end{table*}
\subsection{DiCOVA}

Once again, DiCOVA hosted two COVID-19 audio diagnostic sub-challenges. Both sub-challenge datasets were subsets of the crowd sourced Coswara dataset \cite{SharmaKKRCRGG20}. The first sub-challenge, named Track-1, comprised of a set of 1,274 forced cough audio recordings from 1,274 individuals totalling 1.66 hours. The second, Track-2, was a multi-modality challenge, where 1,199 individuals provided three separate audio recordings; deep breathing, sustained vowel phonation, and counting from 1-20. This dataset represented a total of 14.9 hours of recording. The train-test splits are detailed in Table \ref{tab:DiCOVA}

\begin{table}[t]
  \caption{The results for cross dataset experiments.}
  \label{tab:cross}
  \centering
  \begin{tabular}{ l | c | c | c }
    \toprule
     & \multicolumn{3}{c}{Test Set}\\
    \cline{2-4}
    Train Set & DiCOVA & ComParE & COUGHVID \\
    \hline
    DiCOVA & .799 & .554 & .464\\
    \hline
    ComParE & .512 & .732 & .552\\
    \hline
    EPFL & .395 & .518 & .566\\
    \hline
    All & .673 & .717 & .531\\
    \bottomrule
  \end{tabular}
\end{table}

\section{Results \& Discussion}

The results from the array of experiments with CIdeR and the 3 baseline models are detailed in Table~\ref{tab:results}. CIdeR performed strongly across all four sub-challenges, achieving AUCs of 0.799 and 0.787 in the DiCOVA Track 1 and 2 sub-challenges, respectively, and 0.732 and 0.787 in the ComParE CCS and CSS sub-challenges. In the DiCOVA cough sub-challenge, CIdeR significantly outperformed all three baseline models based on 95\,\% confidence intervals calculated following\cite{hanley_mcneil_1982}, and in the DiCOVA breathing and speech sub-challenge it achieved a higher AUC although the improvement over the baselines was not significant. Conversely, while CIdeR performed significantly better than all three baseline models in the ComParE speech sub-challenge based on 95\,\% confidence intervals calculated following\cite{hanley_mcneil_1982}, it performed no better than baseline in the COMPARE cough sub-challenge. One can speculate that this may have resulted from the small dataset sizes favouring the more classical machine learning approaches which do not need as much training data.

\subsection{Limitations} A key limitation with both the ComParE and DICOVA COVID challenges is the size of the datasets.  Both datasets contain very few COVID-positive participants. Therefore, the certainty in results is limited and this is reflected in the large 95\,\% confidence intervals detailed in Table~\ref{tab:results}. This issue is compounded by the demographics of the datasets. As detailed in \cite{brown2020exploring} and in \cite{muguli2021dicova} for the ComParE datasets and the DiCOVA datasets, respectively, not all demographics from society are represented evenly -- most notably, there is poor coverage of age and ethnicity and both datasets are skewed towards the male gender.

In addition, the crowd-sourced nature of the datasets introduces some confounding variables. Audio is a tricky sense to control. It contains a lot of information about the surrounding environment. As both datasets were crowd-sourced, there could have been correlations between ambient sounds and COVID-19 status, for example, sounds characteristic of hospitals or intensive care units being more often present for COVID-19-positive recordings compared to COVID-19-negative recordings. As the ground truth labels for both datasets were self reported, presumably the participants knew at the time of recording whether they had COVID-19 or not. One could postulate that the individuals who knew they were COVID-19-positive might have been more fearful than COVID-19-negative participants at the time of recording, an audio characteristic known to be identifiable by machine learning models \cite{GeorgeEmotion}. Therefore, the audio features which have been identified by the model may not be specific audio biomarkers for the disease.

We note that both the DiCOVA Track 1 and ComParE CCS sub-challenges were cough recordings. Therefore, there was an opportunity to utilise both training sets. Despite having access to both the DiCOVA and ComParE datasets, training on the two datasets together did not yield a better performance on either of the challenges' test sets. Additionally, a model which performed well on one of the challenges test sets would see a marked drop in performance on the other challenge's test set. We run cross dataset experiments to analyse this effect further. For these experiments, we also included the COUGHVID dataset\cite{orlandic2020coughvid} in which COVID-19 labels were assigned by experts and not as a results of clinically validated test. The results in Table~\ref{tab:cross} show that the trained models for each dataset do not generalise well and perform poorly on excluded datasets. This is a worrying find, as it suggests that audio markers which are useful in COVID classification in one dataset are not useful or present in the other dataset. This agrees with the concerns presented in \cite{coppock2021SevenGrainsofSalt} that current COVID-19 audio datasets are plagued with bias, allowing for machine learning models to infer COVID-19 status, not by audio biomarkers uniquely produced by COVID-19, but by other correlations in the dataset such as nationality, comorbidity and background noise.

\textbf{\textit{Future Work}} One of the most important next steps is to collect and evaluate machine learning COVID-19 classification on a larger dataset that is more representative of the population. To achieve optimal ground truth, audio recordings should be collected at the time that the Polymerase Chain Reaction (PCR) test is taken, before the result is known. This would ensure full blinding of the participant to their COVID status and exclude any environmental audio biasing in the dataset. The Cycle Threshold (CT) of the PCR test should also be recorded, CT correlates with viral load \cite{singct} and therefore would enable researchers to determine the model's classification performance to the disease at varying viral loads. This relationship is critical in assessing the usefulness of any model in the context of a mass testing scheme, since the ideal model would detect a viral load lower than the level that confers infectiousness \cite{SAGE, Peto2021.01.13.21249563}. Finally, studies similar to \cite{bartlpokorny2020voice}, directly comparing acoustic features of COVID-positive and COVID-negative participants should be conducted on all publicly available datasets.

\section{Conclusion}
Cross-running CIdeR on the two 2021 Interspeech COVID-19  diagnosis from cough and speech audio challenges has demonstrated the model's adaptability across multiple modalities. With little modification, CIdeR achieves competitive results in all challenges, advocating the use of end-2-end deep learning models for audio processing thanks to their flexibilty and strong performance.

\section{Funding}

The support of the EPSRC Center for Doctoral Training in High Performance Embedded and Distributed Systems (HiPEDS, Grant Reference EP/L016796/1) is gratefully acknowledged along with the UKRI CDT in Safe \& Trusted AI. The authors further acknowledge funding from the DFG (German Research Foundation) Reinhart Koselleck-Project AUDI0NOMOUS (grant agreement No.\ 442218748) and the Imperial College London Teaching Scholarship.

\bibliographystyle{IEEEtran}

\bibliography{template}


\end{document}